\documentclass[10pt]{llncs}
\usepackage[letterpaper, margin=.8in]{geometry}
\usepackage{amsmath,amsfonts}
\usepackage[affil-it]{authblk}
\usepackage{amstext}
\usepackage{amsgen}
\usepackage{amsbsy}
\usepackage{amsopn}
\usepackage{epsfig}
\usepackage[ampersand]{easylist}
\usepackage{multicol,lipsum,graphicx,float}
\usepackage[english]{babel}
\usepackage[ampersand]{easylist}
\newtheorem{theorema}{Theorem}
\newtheorem{algorithm}[theorema]{Algorithm}

\newcommand{\Prob}{\mathcal{P}}
\newcommand{\msp}{\hspace{0.5cm}}
\newcommand{\LSAT}{$\mathbb{L}$$(${\rm SAT}$)$}

\def\IMAGESPATH{.}

\begin{document}

\title{The fast parallel algorithm for CNF SAT without algebra}
\author{\textbf{Carlos Barr\'{o}n-Romero}\\
 cbarron@correo.azc.uam.mx }
\institute{ Universidad Aut\'{o}noma Metropolitana, Unidad
 Azcapotzalco, M\'{e}xico \\
 Av. San Pablo No. 180, Col. Reynosa Tamaulipas,
 C.P. 02200 \\
 Ciudad de M\'{e}xico, M\'{e}xico
}
\date{2017}
\maketitle

\begin{abstract}

A novel modified numerical parallel algorithm  for solving the classical Decision Boolean Satisfiability problem with clauses in conjunctive normal form is depicted. The approach for solving SAT is without using algebra or other computational search strategies such as branch and bound, back-forward, tree representation, etc. The method is based on the special class of problems, Simple Decision Boolean Satisfiability problem. The design of the main algorithm includes parallel execution,
object oriented, and short termination as my previous versions but it keep track of the parallel tested unsatisfactory binary values to improve the efficiency and to favor short termination. The resulting algorithm is linear with respect to the number of clauses plus a process data on the partial solutions of the Simple Decision Boolean Satisfiability problems and it is bounded by $2^{n}$ iterations where $n$ is the number of logical variables. The novelty for the solution is a linear algorithm, such its complexity is less or equal than the algorithms of the state of the art.

\end{abstract}


\begin{multicols}{2}

\section{Introduction}

The modeling for solving the Decision
Boolean Satisfiability problem (SAT) is based in Reducibility (see~\cite{arXiv:Barron2010}, the chapter 6).
This term means the ability to solve a problem by finding and
solving simple subproblems. The algorithm of this work results from applying simple Decision
Boolean Satisfiability Problems for solving any SAT. 

The notation and conventions for SAT are well know: A boolean variable only takes the values: $0$ (false) or $1$ (true). The logical operators are \textbf{not}: $\overline{x};$ \textbf{and}: $\wedge ,$ and \textbf{or}: $\vee.$
Hereafter, $\Sigma =\left\{0,1\right\}$ is the corresponding binary
alphabet and $X$ is the set of logical variables $\{x_{n-1},\ldots,x_0\}$. A binary string , $w$ $\in$ $\Sigma^n$ is mapped to its
corresponding binary number in $[0,2^n-1]$ and reciprocally. Moreover, a normal conjunctive form clause $x_{n-1} \vee \overline{x}_{n-2}\ldots x_1 \vee x_0$ corresponds to the
binary string $b$=$b_{n-1}b_{n-2}\ldots b_{1} b_0$ where $b_{i}=\left\{
\begin{array}{cc}
0 & \text{if }\overline{x}_{i}, \\
1 & \text{otherwise.}%
\end{array}
\right.$ Also, $\overline{b}=\overline{b}_{n-1}\overline{b}_{n-2}\ldots \overline{b}_{1} \overline{b}_0$ where $b_i$ are the digits of $b$.

The problem  is for determining
when a CNF formula $\varphi \in$ \LSAT or $\notin$ \LSAT, where \LSAT $=$ $\{\varphi\,|\, \varphi$ is satisfiable a CNF logical formula $ \}$ or equivalently, there is a witness $w$ in $\{0,1\}^n$ such that $\varphi(w) \equiv 1$ where $n$ is the $\varphi$'s number of logical variables. The main characteristic of SAT is that all its clauses use $n$ or less variables,  but SSAT´s clauses use $n$ variables. The clauses of SAT and SSAT can have repeated and any order, i.e., the variables could be in clauses far away.

The selection of the Conjunctive normal form (CNF) version of SAT is justified for the
logical equivalence and the huge literature. Talking about SAT is
immediately related to NP Class of computational problems and its
algorithms~\cite{Pudlak1998,Zhang:2001:ECD:603095.603153,Zhang2002,TOVEY198485,coe:Woeginger2003},
~\cite{pnp:page,Fortnow:2009:SPV:1562164.1562186,Cook:2000,Gutfreund:2007:NLH:1341675.1341676}.

Classical SAT problems are depicted as CNF (k,n)-satisfiability or (k,n)-SAT, where the number of variables of the problem is $n$ but each clause have only $k$ variables. By example, with $n=7$ an instance of the (3,7)-SAT is $(x_2 \vee \overline{x}_4 \vee x_6)$ $\wedge$ $(\overline{x}_0 \vee x_1 \vee x_2)$ $\wedge$ $(\overline{x}_1 \vee x_3 \vee \overline{x}_4).$

The algorithm depicted here solves any type of CNF SAT, it means that the clauses are in CNF with any number of the $n$ given logical variables. By example, with $n=8$, an arbitrary instance of SAT is $(x_4 \vee \overline{x}_5 \vee x_7)$ $\wedge$$(\overline{x}_2 \vee \overline{x}_4)$ $\wedge$ $(x_0 \vee \overline{x}_1 \vee x_2)$ $\wedge$ $(x_1 \vee \overline{x}_3 \vee x_4 \vee \overline{x}_5).$

Moreover, this paper focuses in my special parallel algorithm for solving any type of SAT without algebra, the previous results are in~\cite{arXiv:Barron2016,comtel2016:Barron2016a}. 

Briefly, the main results in~\cite{arXiv:Barron2016,arXiv:Barron2016b}, ~\cite{comtel2016:Barron2016a} are:
\begin{enumerate}
    \item The search space of the solutions (satisfiable logical values) 
    of the SAT and SSAT is $\Sigma^n.$
    \item A SSAT with $n$ logical variables, and $m$ clauses , it 
    is solved in $\mathbf{O}(1)$. The comparison of $m$ versus $2^n$  
    is a sufficient condition to state the answer (yes or no) without any 
    process and without any witness.      
    \item An instance of SSAT is not satisfiable when it 
    has $m=2^n$ different clauses. This special instances of SSAT are 
    named blocked boards. By example, in $\Sigma$ and $\Sigma^2$:
    \begin{equation*}
\begin{tabular}{|c|}
\hline
$x_{0}$ \\ \hline
1 \\ \hline
0 \\ \hline
\end{tabular}
\ \ \
\begin{tabular}{|c|c|}
\hline
$x_{1}$ & $x_{0}$ \\ \hline
0 & 0 \\ \hline
1 & 1 \\ \hline
0 & 1 \\ \hline
1 & 0 \\ \hline
\end{tabular}%
\end{equation*}

    \item A clause is translated to a binary number $b$
     then $\overline{b}$
    is for sure an unsatisfactory binary number.
    \item The algorithms based in SSAT for solving SAT are computables and they do not
    require complex computational procedures as sorting, matching, back and forward, or algebra.
     
\end{enumerate}

The main changes with respect to the previous version are:

\begin{description}
    \item[Cooperation]  The two algorithms of deterministic search~\ref{alg:SAT_one} and random search~\ref{alg:SAT_two} shares with a new algorithm~\ref{alg:UpdtCand} the tracking of the failure candidates, which it also searches sequentially a satisfactory assignation.
    \item[Intensive parallel] The random search algorithm~\ref{alg:SAT_two} can test $2^p$ candidates in parallel with algorithm~\ref{alg:test Cand} in $2^p$ independent processors.
\end{description}   

The changes in the previous algorithms of ~\cite{arXiv:Barron2016,comtel2016:Barron2016a} are designed for keeping the computability and efficiency besides the cooperation for keeping tracking  the unsatisfactory tested candidates. Also, the minimum interaction does not alter the previous design for short termination or no more than $2^n$ iterations.
%
\begin{figure}[H]
\begin{center}
\psfig{figure=\IMAGESPATH/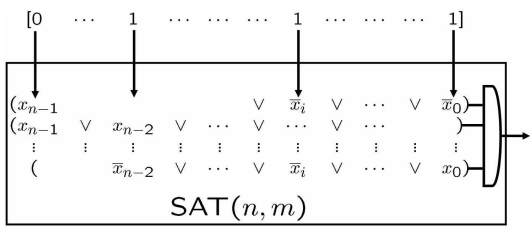,width=70mm}
\caption{SAT as an electronic circuit}~\label{fig:BoxSAT}
\end{center}
\end{figure}

The article is organized as follow. Section~\ref{sc:proper} depicts a selected set of properties and propositions for the parallel algorithm for solving SAT (more details are in~\cite{arXiv:Barron2016,arXiv:Barron2016b,comtel2016:Barron2016a}). The next section~\ref{sc:AlgSAT} depicts the components, design's considerations and the parallel algorithm. Section~\ref{sc:complex_par_algorithm} contains the computational and complex analysis of the parallel algorithm. The next section~\ref{sc:results}and the subsection~\ref{ssc:NotAlgSAT} depict theoretical results about complexity and the consequence of the parallel algorithm for solving SAT. Finally, the last section depicts the conclusions and the future work.

\section{Properties of SAT}~\label{sc:proper}

This section depicts the main propositions used to build the parallel algorithm.

\begin{proposition}
~\label{prop:binNumBlock} Any CNF clause $x$ over the variables
$(x_{n-1},$ $\ldots,$ $x_0)$ corresponds a binary number $b$ $=$
$b_{n-1}b_n$ $\ldots$ $b_0 $. Then
$x(b) \equiv 1$ and $x(\overline{b}) \equiv 0$
where $b$ is the translation of the clause $x$, the values of the boolean variables correspond to
the
binary digits of $b$ and $\overline{b}$

\begin{proof}
Without loss of generality,  $x$ $=$ $x_{n-1}\vee
\overline{x}_{n-2}\vee \ldots \vee \overline{x}_1 \vee x_0$, $b$ $=$ $10\ldots01$ is the translation of $x$, and $\overline{b}$ $=$ $01\ldots10$ is the complement binary number of $b$. Then $x(b)=1 \vee \overline{0}\ldots \vee \overline{0} \vee 1$ $\equiv$ $1\vee 1\ldots\vee 1 \vee 1$ $\equiv$ $1$, and $x(\overline{b})=0\vee \overline{1}\ldots\vee \overline{1}\vee 0$ $\equiv$ $0\vee 0\ldots\vee 0$ $\equiv$ $0.$
\end{proof}
\end{proposition}

The translation of the SSAT's clauses allow to define a table of binary numbers or a board. The following boards with $2^n$ different clauses have not a satisfactory assignation in $\Sigma
$ and $\Sigma ^{2}$:

\begin{equation*}
\begin{tabular}{|c|}
\hline
$x_{0}$ \\ \hline
1 \\ \hline
0 \\ \hline
\end{tabular}
\ \ \
\begin{tabular}{|c|c|}
\hline
$x_{1}$ & $x_{0}$ \\ \hline
0 & 0 \\ \hline
1 & 1 \\ \hline
0 & 1 \\ \hline
1 & 0 \\ \hline
\end{tabular}%
\end{equation*}

These are examples of blocked boards. It is clear by the previous proposition,
that the corresponding SSAT are $\varphi_1=(x_0)\wedge (\overline{x}_0)$ and $\varphi_2=(\overline{x}_1 \vee \overline{x}_0)$ $\wedge$ $(x_1 \vee x_0)$ $\wedge$ $(\overline{x}_1 \vee x_0)$ $\wedge$ $(x_1 \vee \overline{x}_0)$ have not a solution because each binary number has its binary
complement.

A brief justification for not using algebra and getting efficiency for solving SAT is depicted in the following proposition. Doing algebra or branch and bound or tree search cause more operations and iterations than reading the SAT's clauses.

\begin{proposition}~\label{prop:FtFvyFnv} Let $F$ be a logical formula and $v$ is logical variable, which it is not in $F$. Then 

$$
\begin{array}{ccc}
\left( F \right) & \equiv &
\begin{array}{cc}
& \left( F \vee v\right) \\
\wedge & \left( F\vee \overline{v}\right)%
\end{array}%
\end{array}%
$$

\begin{proof} The results follows from the algebraic laws of distribution and factorization $
\left( F \right)  \equiv  \left( F \wedge \left( v\vee
\overline{v}\right) \right).$
\end{proof}
\end{proposition}

Moreover, the previous proposition states a logical reciprocal equivalence transformation between  SAT and
SSAT. By example, let $\varphi_3$ be a SAT$(4,2)$,

$$
\varphi_3= \begin{array}{ccccc}
\ \    & (x_{3}& \vee \ \overline{x}_{2}& & \vee \  x_{0})  \\
\wedge &   &  ( x_{2}& \vee \ x_{1} & \vee \  \overline{x}_{0}).%
\end{array}%
$$

Using the previous proposition, $\varphi_3$ is equivalent to a $\varphi_4$ of type  SSAT$(4,4)$, where:
$$
\varphi_4=\begin{array}{ccccc}
\ \    & (x_{3}& \vee \ \overline{x}_{2}& \vee \ x_1  & \vee \ x_{0}) \\
\wedge & (x_3  & \vee \  \overline{x}_{2}& \vee \ \overline{x}_{1} &
\vee \ x_{0}) \\
\wedge & (x_3  & \vee  x_{2}& \vee \ x_{1} & \vee \ \overline{x}_{0}) \\
\wedge & (\overline{x}_3  & \vee  x_{2}& \vee \ x_{1} & \vee \
\overline{x}_{0})
.%
\end{array}%
$$

Both are equivalent, and they have the same satisfactory assignations, but it is not necessary to use algebra for solving them, it is only necessary to read their clauses.

By inspection and ordering variables and clauses, $\varphi_3$'s satisfactory assignation only need that
 $x_3=1$ and $x_1=1$ (its other variables do not care). 
 On the other hand, $\varphi_4$'s satisfactory assignations are
  $\{ 1010,$ $1011,$ $1110,$ $1111\}$. These satisfactory assignations
 correspond to the variables with $x_3=1$ y $x_1=1.$

This toy example depicts that by doing algebraic procedures it is necessary to apply the laws of factorization and distribution after finding matching between clauses and variables, but in order to find matching between clauses and variables it is necessary to sort or to use searching matching procedures to determine the matching between clauses and variables under an appropriated data structures. A review of the state of the art for the algorithms for SAT, depicts that they use expensive strategies, by example, branch and bound, backtracking, sorting and matching, etc. It means more operations than reading of the clauses as they appear in a problem. The parallel algorithm executes a deterministic search algorithm~\ref{alg:SAT_one} that need to read the clauses s they are plus a data process $\times\,\Theta.$

\section{Algorithms for SAT}~\label{sc:AlgSAT}

There is a bijection between $x \subset X$ and an unique identification number. This bijective function is similar to the given in ~\cite{arXiv:Barron2010}, Prop. 4.

$$
\begin{array}{ccl}
\vbox{\hbox{Set of logical}\hbox{variables}} &  & \mathbb{N} \\
\{\} & \leftrightarrow & 0 \\
\{x_{0}\} & \leftrightarrow & 1=\binom{n}{0} \\
\{x_{1}\} & \leftrightarrow & 2=\binom{n}{0}+1 \\
\vdots & \vdots & \vdots \\
\{x_{1},x_{0}\} & \leftrightarrow & k_1=\sum_{k=0}^{1}\binom{n}{k} \\
\{x_{2},x_{0}\} & \leftrightarrow & k_2=\sum_{k=0}^{1}\binom{n}{k}+1 \\
\vdots & \vdots & \vdots \\
X & \leftrightarrow &
2^{n}-1 = \sum_{k=0}^{n-1}\binom{n}{k}%
\end{array}%
$$

The function  $IdSS:2^{X}\rightarrow \lbrack 0,2^{n}-1]$ is estimated by the following algorithm to get a bijection between a subset of variables of $X$ and its identification number.

\noindent\hrulefill 
\begin{algorithm} ~\label{alg:IdxSet} Unique identification of SSAT by its variables

\noindent\textbf{Input:} $x=\{x_k,\ldots,x_1,x_0\}$: set of variables indexed in  $[0,n]$ and in descending order.

\noindent \textbf{Output:} $ix$: integer value;

\noindent // identification number in $[0,2^n-1]$ for the indices in the set $x$.

\noindent \textbf{Memory}: $base$: integer; $v,t$: (k+1)-ism array structure of indices as number where its digits are in numerical base $n$, i.e., its digits are $\{n-1,n-2,\ldots,1,0\}$

\noindent\hrulefill
\begin{easylist}
\renewcommand{\labelitemi}{\ }

\item $ix = 0;$

\item $k = |x|;$ // $|\cdot|$ cardinality of a set.

\item \textbf{if} $(k$ == $0)$ \textbf{then}

\item \msp \textbf{output}: \text{``}$ix$";

\item \msp \textbf{return};

\item \textbf{end if}

\item $base = 0$;

\textbf{for} $j=0,k-1$ \textbf{do}

\item \msp $base = base + \binom{n}{j};$ 

\item \msp // $\binom{\cdot}{\cdot}$
binomial or Newton coefficient  

\item \textbf{end for}

\item \textbf{concatenate} $v=\{v_k,\ldots,v_0\}$; 

\item // array of size $k+1$ as number for the indices in descending order

\item \textbf{while} $(indices(x) < (indices(v))$ \textbf{do}.

\item \msp $t=v;$

\item \msp \textbf{while} (1) {\bf do}

\item \msp \msp $t= incrementa(t,1);$

\item \msp \msp // increase by one the $t$'s indices

\item \msp \msp as a number in the numerical base $n$

\item \msp \msp \textbf{if} $indices(t)$ are different
 \item \msp \msp {\bf and} in descending order \textbf{then}

\item \msp\msp\msp \textbf{break}
\item \msp\msp\msp// $t$ is a valid set of different descent
\item \msp\msp\msp//ordering indices 

\item \msp\msp \textbf{end if}

\item \msp \textbf{end while};

\item \msp $v=t$;

\item \msp $ix=ix+1$;

\item \textbf{end while} 

\item // this loop end when the index set $x$ is founded

\item \textbf{output}: \text{``}$base+ix$";

\item \textbf{return};

\end{easylist}
\noindent\hrulefill
\end{algorithm}

The next algorithms contain the changes for cooperation and intensive parallel with the following algorithm for registering the failed candidates.

\noindent\hrulefill 
\begin{algorithm}
~\label{alg:UpdtCand} Updated failed candidate

\noindent \textbf{Input:} $n:$ number of variables and $\varphi:$ SAT formula;

\noindent \textbf{Messages}c: binary number in $\Sigma^n$; 

\noindent\textbf{Do:} get\_in(c, L\_C: List);

\noindent \textbf{Output}: Nothing or message and short termination.

\noindent \textbf{Memory}:

L\_c : List of binary numbers; 

\noindent \textbf{Exclusive memory}:

n\_cand$:=2^n$ : integer; 

cand\_stat$[0,2^{n-1}] := 1$: circular list of boolean, next,prior: integer pointers; // 1: viable, 0: failure for satisfying $\varphi$

next\_c:=0: Integer pointer to the next available cand\_stat;

\noindent\hrulefill
\begin{easylist}
\renewcommand{\labelitemi}{\ }

\item {\bf while} $(1)$ {\bf do}

\item \msp {\bf while} not empty $($L\_c$)$ {\bf do}

\item \msp  \msp $c$ $:=$ get\_out$($ L\_c $)$;

\item \msp \msp \textbf{if} (cand\_sta$[c] == 0$) \textbf{then}

\item \msp \msp \msp \textbf{continue};

\item \msp \msp \textbf{end if}

\item \msp  \msp  cand\_sta$[c] := 0$;

\item \msp  \msp n\_cand := n\_cand - 1;

\item \msp \msp \textbf{if} $($n\_cand == $0)$ \textbf{%
then}

\item \msp \msp \msp \textbf{output}: \text{``}The algorithm~\ref{alg:UpdtCand} confirms 

\item \msp \msp \msp $\varphi\notin$\LSAT after reviewing $\Sigma^n$.";

\item \msp \msp \msp \textbf{stop all};

\item \msp \msp \textbf{end if}

\item \msp \msp Update\_Circular\_List(cand\_stat, $c$, next\_c);

\item \msp {\bf end while}

\item \msp \textbf{if} $($n\_cand $>$ $0)$ \textbf{%
then}

\item \msp  \msp $c$ $:=$ next\_c;

\item \msp \msp \textbf{if} $(\varphi(c) == 1)$ \textbf{%
then}

\item \msp \msp \msp \textbf{output}: \text{``}The 
algorithm~\ref{alg:UpdtCand} confirms 

\item \msp \msp \msp$\varphi\in$\LSAT,

\item \msp \msp \msp$c$ is a satisfactory assignation.";

\item \msp \msp \msp {\bf stop all};

\item \msp \msp {\bf end if}

\item \msp \msp  cand\_sta$[c] := 0$;

\item \msp \msp  n\_cand := n\_cand - 1;

\item \msp \msp \textbf{if} $($n\_cand == $0)$ \textbf{%
then}

\item \msp \msp \msp \textbf{output}: \text{``}The algorithm~\ref{alg:UpdtCand} confirms 

\item \msp \msp \msp $\varphi\notin$\LSAT after reviewing $\Sigma^n$.";

\item \msp \msp \msp \textbf{stop all};

\item \msp \msp \textbf{end if}

\item \msp \msp Update\_Circular\_List(cand\_stat, $c$, next\_c);

\item \msp {\bf end if}

\item {\bf end while}

\end{easylist}
\noindent\hrulefill
\end{algorithm}

The integer pointers and the next available candidate of the circular list cand\_stat is updated by the routine Update\_Circular\_List. There is not dynamic memory, just integer pointers update.

\noindent\hrulefill 
\begin{algorithm}
~\label{alg:NumSigma} Number$_\Sigma$

\noindent \textbf{Input:} (rw: set of integer values, for the identification of the logical variables, k: set of binary values of each variable of rw)

\noindent \textbf{Output}: ($k_\Sigma$: array binary value in $\Sigma^n$).

\noindent \textbf{memory}:

$i$ : integer; 

\noindent\hrulefill
\begin{easylist}
\renewcommand{\labelitemi}{\ }

\item \textbf{for} i$:=n-1$ {\bf downto} $0$ {\bf do}

\item \msp {\bf if} $(i  \in$ rw $)$ {\bf then}

\item \msp \msp $k_\Sigma[i]$ $:=$ {\bf value in} $k$ {\bf of the variable} $i$;

\item \msp {\bf else}
\item \msp \msp $k_\Sigma[i]$ $:=$ random selection from $\Sigma$;

\item \msp {\bf end if}

\item {\bf end for}

\item {\bf return} $k_\Sigma$;

\end{easylist}
\noindent\hrulefill
\end{algorithm}

\noindent\hrulefill 
\begin{algorithm}
~\label{alg:UpdtSSAT} Updated SSAT

\noindent \textbf{Input:} (SSAT: List of objects, rw: clause).

\noindent \textbf{Output}: SSAT: Updated list of objects for each
SSAT with  identification $IdSS(r)$.

\noindent Each SSAT$(IdSS(r))$ updates its $S:$ list of the solutions,
where $S[0:2^{n-1}]$: is an array as a doble chained list of integer , $previous$, $next$ : integer;

\noindent \textbf{Memory}:
$ct:=0$ : integer; $k, k$\_aux$:$ integer;
$first:=0$: integer; $last=2^{n}-1$: integer;

\noindent\hrulefill
\begin{easylist}
\renewcommand{\labelitemi}{\ }

\item \textbf{if} $($ SSAT$(IdSS(rw))$ not exist$)$ \textbf{then}

\item \msp {\bf build object} SSAT$(IdSS(rw))$

\item \textbf{end if}

\item \textbf{with} SSAT$(IdSS(rw))$

\item \msp $k$ $:=$ \textbf{clause to binary} ($rw$);

\item \msp $k$\_aux := $k$;
\item \msp {\bf if}  $($size$(k)$ $< \varphi.n)$ {\bf then}
\item \msp \msp $k$\_aux := number$_\Sigma$(set\_of\_variables$(rw),k)$;
\item \msp {\bf end if}

\item \msp \textbf{if} $(\varphi(k$\_aux$) == 1)$ \textbf{%
then}

\item \msp \msp \textbf{output}: \text{``}The 
algorithm~\ref{alg:UpdtSSAT} confirms 

\item \msp \msp$\varphi\in$\LSAT,

\item \msp \msp$k$\_aux is a satisfactory assignation.";

\item \msp \msp {\bf stop all};

\item \msp {\bf else}

\item \msp \msp Updated\_failed\_candidate($k$\_aux); // algorithm~\ref{alg:UpdtCand}

\item \msp \textbf{end if}

\item \msp // $\overline{k}$ of size $n$ is not a satisfactory number for $\varphi$. 

\item \msp // See proposition~\ref{prop:binNumBlock}.

\item \msp \textbf{if} $($size$(k) == \varphi.n)$

\item \msp \msp Updated\_failed\_candidate$(\overline{k})$; // algorithm~\ref{alg:UpdtCand}

\item \msp \textbf{end if}

\item \msp \textbf{if} $(S[\overline{k}].previous \neq$
$-1)$

\item \msp  \textbf{\ or } $(S[\overline{k]}.next \neq$
$-1)$ \textbf{then}

\item \msp \qquad $S[S[\overline{k}].previous].next$ $:=$ 
$S[\overline{k}].next$;

\item \qquad \qquad  $S[S[\overline{k}].next].previous$ $=$ 
$S[\overline{k}].previous$;

\item \qquad \qquad \textbf{if} $(\overline{k}$ $==$ $first)$
\textbf{then}

\item \qquad \qquad \qquad $first$ $:=$ $S[\overline{k}].next$;

\item \qquad \qquad \textbf{end if}

\item \qquad \qquad \textbf{if} $(\overline{k}$ $=$ $last)$
\textbf{then}

\item \qquad \qquad \qquad $last$ $:=$ $S[\overline{k}].previous$;

\item \qquad \qquad \textbf{end if}

\item \qquad \qquad $S[\overline{k}].next :=-1$;

\item \qquad \qquad $S[\overline{k}].previous :=-1;$

\item \msp \qquad $ct:=ct+1;$

\item \msp \textbf{end if}

\item \msp \textbf{if} $(ct$ = $2^{n})$ \textbf{then}

\item \msp \msp \textbf{output:} \text{``} The 
algorithm~\ref{alg:UpdtSSAT} confirms 

\item \msp \msp$\varphi\notin$\LSAT

\item \msp \msp SSAT$(IdSS(rw))$ is a blocked board.";

\item \msp \msp \textbf{stop all};

\item \msp \textbf{end if}

\item \textbf{end with}

\item \textbf{return}

\end{easylist}
\noindent\hrulefill
\end{algorithm}

\noindent\hrulefill 
\begin{algorithm}
~\label{alg:SAT_one} Solve $\varphi\in$\LSAT.

\noindent\textbf{input:} $n:$ number of variables and $\varphi:$ SAT formula;

\noindent \textbf{output:} Message and if there is a solution the witness $x$, such that
$\varphi(x)=1$.

\noindent \textbf{Memory}: 

$r$: set of variables de $X$;

SSAT=null: List de objets SSAT.

\noindent\hrulefill
\begin{easylist}
\renewcommand{\labelitemi}{\ }

\item \textbf{while} not(eof\_clauses($\varphi$));

\item \msp $r=$ $\varphi$.read\_clause;

\item \msp algorithm.\ref{alg:UpdtSSAT} Updated SSAT(
SSAT,r).

\item \textbf{end while};

\item \textbf{with list} SSAT

\item \msp  // $\times\theta$ cross product and natural joint

 \item \msp
\textbf{estimate} $\Theta$ = $\times\theta$ with all SSAT$(IdSS(r))$;

\item \msp \textbf{if} $\Theta$ = $\emptyset$
\textbf{then}

\item \msp \msp \textbf{output}: \text{``}The
algorithm~\ref{alg:SAT_one} confirms 

\item \msp \msp $\varphi\notin$\LSAT.

\item \msp\msp The
SSAT$(IdSS(r))$

\item \msp\msp are incompatible."

\item \msp\msp\textbf{stop all};

\item \msp \textbf{otherwise}

\item \msp \msp \textbf{output}: \text{``}The
algorithm~\ref{alg:SAT_one} confirms 

\item \msp \msp $\varphi\in$\LSAT.

\item \msp \msp $s$ 
 is a satisfactory assignation, $s\in\Theta$.

\item \msp \msp The
SSAT$(IdSS(r))$

\item \msp \msp are compatible.

\item \msp \msp \textbf{stop all};

\item \msp \textbf{end with}

\item \textbf{end if}

\end{easylist}
\noindent\hrulefill
\end{algorithm}

\noindent\hrulefill 
\begin{algorithm}
~\label{alg:test Cand} Test $\varphi(\cdot)$ 

// It evaluates and updates failure candidates.

\noindent\textbf{Input:} $c$: integer value in $\Sigma^n$.

\noindent\textbf{Output:}  none.

\noindent\hrulefill
\begin{easylist}
\renewcommand{\labelitemi}{\ }

\item \textbf{if} $($Cand\_stat$(c)$ $==$ $1)$ \textbf{then}

\item \msp \textbf{if} $(\varphi(c)$ == $1)$ \textbf{then}

\item \msp \msp \textbf{output}: \text{``}The
algorithms~\ref{alg:test Cand} and~\ref{alg:SAT_two} confirm 

\item \msp \msp $\varphi\in$\LSAT.

\item \msp \msp $c$ is a satisfactory assignation.";

\item \msp \msp \textbf{stop all};

\item \msp {\bf else}

\item \msp \msp Updated\_failed\_candidate$(c)$; // algorithm~\ref{alg:UpdtCand}

\item \msp \textbf{end if}

\item \textbf{end if}

\item \textbf{return}

\end{easylist}
\noindent\hrulefill
\end{algorithm}

The next algorithm is a modified version of the algorithm 4
in~\cite{arXiv:Barron2015b}. The selected candidates are unique and randomly selected from $\Sigma^n$ interpreted as the interval of natural numbers $[0,2^n-1].$

\noindent\hrulefill 
\begin{algorithm}
~\label{alg:SAT_two} Solve $\varphi\in$\LSAT  by a random search of $[0,2^n-1].$

\noindent\textbf{Input:} $n:$ number of variables and $\varphi:$ SAT formula;

\noindent\textbf{Output:}  Message and if there is a solution, the witness $s$
($s\in[0,2^n-1]$), such that $\varphi(s)=1$.

\noindent\textbf{Memory}: $T[0: 2^{n-1}-1]$=$[0: 2^{n}-1]$: integer; $Mi$%
=$2^{n}-1$: integer; $rdm,a$: integer.

\noindent\hrulefill
\begin{easylist}
\renewcommand{\labelitemi}{\ }

\item \textbf{for} i:=0 to $2^{n-1}-2$

\item \msp \textbf{if} $T[i]$ = $i$ \textbf{then}

\item \msp \msp // random selection $rdm \in
[i+1, 2^{n-1}-1]$;

\item \msp \msp $rdm$ $=$ floor($rand()$
$(Mi-i+1.5)$) + $(i+1)$;

\item \msp \msp // rand() return a random number in
 (0,1);

\item \msp \msp // floor(x) integer less than $x$

\item \msp\msp $a$ $=$ $T[rdm]$;

\item \msp\msp $T[rdm]$ $=$ $T[i]$;

\item \msp \msp $T[i]$ $=$ $a$;

\item \msp \textbf{end if}

\item \msp {\bf parallel execution}

\item \msp \msp // Concatenation of all the strings  $\Sigma^p$ 
\item \msp \msp // and the random string from $\Sigma^{n-p}$

\item \msp \msp test$(0T[i])$ by algorithm~\ref{alg:test Cand}; 
\item \msp \msp test$(1T[i])$ by algorithm~\ref{alg:test Cand};

\item \msp {\bf end parallel execution}

\item \textbf{end for}


\item \msp {\bf parallel execution}

\item \msp \msp // Concatenation of all the strings  $\Sigma^p$ 
\item \msp \msp // and the random string from $\Sigma^{n-p}$

\item \msp \msp test$(0T[2^{n-1}-1])$ by algorithm~\ref{alg:test Cand}; 
\item \msp \msp test$(1T[2^{n-1}-1])$ by algorithm~\ref{alg:test Cand};

\item \msp {\bf end parallel execution}

\item \textbf{output}: \text{``}The algorithm~\ref{alg:SAT_two}
confirms $\varphi\notin$\LSAT

\item  after exploring all $\Sigma^n$.";

\item  \textbf{stop all};

\end{easylist}
\noindent\hrulefill
\end{algorithm}

The selection of $2^p$ processors is for simplified the segmentation of the space $\Sigma^n$ in the parallel execution. The upper limit for the iterations of previous algorithm is
$2^{n-1}$, but the number of tested candidates remains $2^n$. It is $\frac{2^n}{2}=2^{n-1}$ by parallel execution by two independent processors of the algorithm~\ref{alg:test Cand}. It is possible to get down the number of iterations by using four independent processor, then number of iterations is $2^{n-2}$.  The simultaneous candidates for testing are $00x$,
$01x$, $10x$ and $11x$ where
$x\in[0,2^{n-2}-1]$. The candidates space $\Sigma^n$ is exploring in each iteration by all the strings of $\Sigma^p$ and one random string from $\Sigma^{n-p}$.
In general with $2^p$ independent processors the upper limit of iterations is $\frac{2^n}{2^p}=2^{n-p}$ but it is not as good as it sounds, see proposition~\ref{prop:parallel_randseaarch}. Also, the randomness is affected by the strings of $\Sigma^p$.

The algorithms~\ref{alg:SAT_one} and~\ref{alg:SAT_two} have short termination when a satisfactory assignation or a blocked board are found, no matters if the number of clauses is huge or the clauses are in disorder and repeated. Also, they send their failed candidates to the algorithm~\ref{alg:UpdtCand}. This causes that the space of candidates of it approximately decreases by a factor of $2^{p+1}$ in each iteration (the number of messages send to it).

\noindent\hrulefill 
\begin{algorithm} Parallel algorithm for SAT
~\label{alg:Parallel}

\noindent\textbf{Input:} $n:$ number of variables and $\varphi:$ SAT formula;

\noindent\textbf{Output}: Message if $\varphi\in$\LSAT or not.

\noindent\hrulefill
\begin{easylist}
\renewcommand{\labelitemi}{\ }

\item \textbf{parallel execution}

\item \msp algorithm~\ref{alg:UpdtCand}($n$, $\varphi$);

\item \msp algorithm~\ref{alg:SAT_one}($n$, $\varphi$);

\item \msp algorithm~\ref{alg:SAT_two}($n$, $\varphi$);

\item \textbf{end parallel execution}

\end{easylist}
\noindent\hrulefill
\end{algorithm}

There are two main ideas for keeping efficient the algorithm~\ref{alg:SAT_two}: 1) a random search using a permutation of the research space $[0,2^n-1],$ and 2) the permutation is computed at the same time of the trials. More information in~\cite{arXiv:Barron2015b,arXiv:Barron2016,comtel2016:Barron2016a}.

The 
algorithm~\ref{alg:SAT_one} behaves as one pass compiler over the reading of the clauses as they are. It does not need more operations in order to pull apart the binary number of the complement of the clause's translation to a binary number from the research space of each SSAT, this allows to detect if a SSAT is a blocked board for short termination. Otherwise, when it finishes to read the clauses
of the given formula $\varphi$, it has all the satisfactory assignations for each 
SSAT of $\varphi$. The final step of the algorithm~\ref{alg:SAT_one}
is to execute with the satisfactory assignations of each SSAT a cross theta joint operation $\times
\theta$. This operation is similar to the cross product and natural theta joint in the relational databases. The definition of  $\times
\theta$ with two set of logical variables  $r$ and $r^\prime$, and with their satisfactory assignations SSAT$(\cdot).S$ is:

SSAT$(IdSS(r))$ $\times\, \theta $ SSAT$(IdSS(r^{\prime }))=$
\begin{enumerate}

\item \textbf{if} $r \cap r^{\prime }=\emptyset$ \textbf{then}

SSAT$(IdSS(r)).S\times \text{SSAT}(IdSS(r^{\prime })).S$.

\item \textbf{if} $r \cap r^{\prime} \ne \emptyset$ \textbf{and} there are common values between 
SSAT$(IdSS(r)).S\text{ and SSAT}(IdSS(r^{\prime })).S$

for variables in $r \cap r^{\prime}$ \textbf{then}

SSAT$(IdSS(r)).S \,  \theta _{r \cap
r^{\prime}}\,$SSAT$(IdSS(r^{\prime })).S$

\item \textbf{if} $ r \cap r^{\prime} \ne \emptyset$  \textbf{and}

there are not common values between 

SSAT$(IdSS(r)).S$  y SSAT$(IdSS(r^{\prime })).S$

for the variables in  $r \cap r^{\prime}$ \textbf{then}
$\emptyset.$

\end{enumerate}

For 1) y 2)  SSAT$(IdSS(r)$ and SSAT$(IdSS(r^{\prime })$ are compatible. For 3) they are incompatible, i.e., there is not a satisfactory assignation for both.

An example of 1) is $\varphi_5=$SAT$\left( 4,5\right)$ with 
$ (x_{3}\vee \overline{x}_{2}) \wedge  (x_{3}\vee x_{2}) \wedge
(\overline{x}_{3}\vee x_{2}) \wedge  (\overline{x}_{1}\vee
\overline{x}_{0}) \wedge  \text{ } (x_{1}\vee x_{0}).$ It has a 
SSAT$\left( 2,3\right) $ for its first three clauses with solutions $x_{3}=1, x_{2}=1.$ It has a SSAT$\left( 2,2\right)
$ from its two last clauses with solutions
$\{(x_{1}, x_{0}) \, | \ (0,1) \vee (1,  0) \}.$ Then the satisfactory assignation for $\varphi_5$ are $\left\{ \left( 1,1\right) \right\}
\times \left\{ \left( 0,1\right) ,\left( 1,0\right) \right\}$ $=$ $\{
(x_{3}, x_{2},x_{1}, x_{0}) \, | \, (1, 1, 0,  1) \vee (1, 1, 1,
0)\}.$

An example of 2) is $\varphi_6=$SAT$\left( 4,5\right)$ $=$
 $(x_{3}\vee \overline{x}_{2}) \wedge
 (\overline{x}_{3}\vee \overline{x}_{2})
\wedge  (\overline{x}_{3}\vee x_{2}) \wedge  (x_{2}\vee
\overline{x}_{1}\vee x_{0}) \wedge  (x_{2}\vee  x_{1}\vee
\overline{x}_{0}).$ It has a SSAT$\left( 2,3\right) $ from its three first clauses with solutions $\{ (x_{3}, x_{2})
\,|\, (0, 0) \} .$ It has a SSAT$\left( 3,2\right) $ from its last two clauses with solutions $\{ (x_{2},
x_{1},  x_{0}) \,|\, (1,  0,   0) \vee (1, 0,  1) \vee (1, 1,  0) \vee
(1,  1, 1) \vee (0, 0,  0) \vee (0,  1, 1) \}.$ Then $\varphi_6
\in$\LSAT, because there are satisfactory assignations for the common value $0$ of the common variable $x_{2}$. The satisfactory assignation for $\varphi_6$ are $\{ (x_{3},  x_{2},  x_{1},  x_{0}) \,| \, (0,  0,
0, 0) \vee (0,  0,  1, 1) \}.$

An example of 3) is $\varphi_7=$SAT$\left( 4,7\right)$ $=$ 
$(x_{3}\vee \overline{x}_{2}) \wedge  (\overline{x}_{3}\vee
\overline{x}_{2}) \wedge (\overline{x}_{3}\vee x_{2}) \wedge
(x_{2}\vee \overline{x}_{1}\vee \overline{x}_{0}) \wedge
(x_{2}\vee \overline{x}_{1}\vee x_{0}) \wedge (x_{2}\vee
x_{1}\vee \overline{x}_{0})
\wedge (x_{2}\vee x_{1}\vee x_{0})%
.$ It has a  SSAT$\left( 2,3\right) $ from its three first clauses with solutions $\{ (x_{3},  x_{2}) \, | \,
(0,  0) \} .$ It has a SSAT$\left( 3,4\right) $ from its last four clauses with solutions $\{ (x_{2},
x_{1}, x_{0}) \,| \, (1,  0,  0) \vee (1, 0,  1) \vee (1,  1, 0) \vee (
1, 1, 1) \}.$ Then $\varphi_7 \notin$\LSAT, because there are not satisfactory assignations to build with the common variable $x_{2}$.

\section{Computability and complex analysis of the parallel algorithm}~\label{sc:complex_par_algorithm}

There are three
properties of the model of computation of the nowadays 
computers: 1) the concept of discrete states; 2) the manipulation of its memory, and 3) its finite alphabet.

The communication between computers as in persons has unsolved issues. Secure communication without eye drooping is one issue and checking and verifying send and receive messages is also an open problem. Here, no protocol for communications is assumed because the time cost could cause an increase of the complexity or not computability at all (forever loop) with noise lines, repeated messages, and collisions. A simple communication in one direction from sender to receiver without acknowledgement from algorithms~\ref{alg:SAT_one} and~\ref{alg:SAT_two} to algorithm~\ref{alg:UpdtCand} is assumed, if there are collisions or losing messages do not affect the time or computability.

Figure~\ref{fig:BoxSAT} depicts a SAT as an electronic circuit of and or gates where the lines are the logical variables. I assume that the evaluation of $\varphi$ is without a computer program, it is a given electronic circuit as a closed box that it works when the signals of the values of the logical variables in its lines are given. The electrons travel into the lines at the speed of light and there is an small time required by the gates for an stable output. Therefore the time cost for evaluating $\varphi$ is a small constant. It is like a city that turns on its lights when the dark comes. There are wonderful videos of cities turn on their lights.

On the other hand, for analyzing the clauses of $\varphi$ it is necessary states and memory to keep the status and information of the reviewing or reading $\varphi$'a clauses. A deterministic automate is not enough for this task, because it is necessary to store and retrieve data about the reviewing process. Therefore an appropriate computational model is a Turing machine.

Assuming appropriate Turing machines for executing the algorithms, the following propositions depicts the computability and complexity of:
\begin{enumerate}
    \item The deterministic search algorithm~\ref{alg:SAT_one} and its data processing of the operator $\times\theta$. 
    \item The random search algorithm~\ref{alg:SAT_two} with $2^p$ processors for exploring $\Sigma^n$ in parallel.
    \item The updated failed candidates algorithm~\ref{alg:UpdtCand} 
    interaction with the algorithms~\ref{alg:SAT_one} and~\ref{alg:SAT_two}, and 
    its sequential search for a satisfactory candidate.
\end{enumerate}

\begin{proposition}
~\label{prop:CompSATOne} The algorithm~\ref{alg:SAT_one} is computable and its iterations are limited by $m$, the $\varphi$'s number of clauses plus the iterations of the operator $\times\,\Theta$.

\begin{proof}

The algorithm executes the algorithms~\ref{alg:IdxSet}, ~\ref{alg:UpdtSSAT} and~\ref{alg:NumSigma}, and the operation $\times\,\Theta$, which can be duplicated by appropriate Turing machines because they correspond to computable operations as sum, product, copy, identify, and fill. 

The operator $\times\,\Theta$ works as the relational data base operator cross product and natural joint. For the set of variables $r$ and $r^\prime$ $\subset X$ and the satisfactory assignations SSAT$(\cdot).S$, 
 SSAT$($Id\_SSAT$(r))$ $\times\, \theta $ SSAT$($Id\_SSAT$(r^{\prime }))=$
\vspace{-2mm}
\begin{enumerate}

\item \textbf{if} $r \cap r^{\prime }=\emptyset$ \textbf{then}

SSAT$($Id\_SSAT$(r)).S\times \text{SSAT}($Id\_SSAT$(r^{\prime })).S$.

\item \textbf{if} $r \cap r^{\prime} \ne \emptyset$ \textbf{and} there are common values between   
SSAT$($Id\_SSAT$(r)).S\text{ and SSAT}($Id\_SSAT$(r^{\prime })).S$

for the variables in  $r \cap r^{\prime}$ \textbf{then}

SSAT$($Id\_SSAT$(r)).S \,  \theta _{r \cap
r^{\prime}}\,$SSAT$($Id\_SSAT$(r^{\prime })).S$

\item \textbf{if} $ r \cap r^{\prime} \ne \emptyset$ \textbf{and}

there are not common values between   

SSAT$($Id\_SSAT$(r)).S$  y SSAT$($Id\_SSAT$(r^{\prime })).S$

for the variables in  $r \cap r^{\prime}$ \textbf{then}
$\emptyset.$
\end{enumerate}

For 1) and 2) the solutions of SSAT$(\cdot).S$ are compatibles and any is a satisfactory assignation. For 3) the solutions of SSAT$(\cdot).S$ are incompatibles, the satisfactory assignation is the empty set.

The algorithm~\ref{alg:SAT_one} always finishes with the solution of the yes-no question: $\varphi \in$\LSAT?. This means a satisfactory assignation or a blocked board is founded, this include the operation $\varphi(\cdot)$ that it estimates one satisfactory assignation when the SSAT's data are compatibles or not solution when SSAT's data are incompatibles.

The failed candidates are sent to the algorithm~\ref{alg:UpdtCand} without a communication protocol. It does not affect the efficiency and the computability of the algorithm~\ref{alg:SAT_one}.

Finally, the iterations correspond to the $\varphi$'s number of clauses and the operator $\times\,\Theta$.
\end{proof}
\end{proposition}

The algorithm~\ref{alg:SAT_one} behaves as one pass compiler with short termination. It requires to read $\varphi$'s clauses plus the conciliation of the solutions of the SSAT subproblems. Examples of the operator $\times\,\Theta$ are:

For 1) $\varphi_5=$SAT$\left( 4,5\right)$ with 
$(x_{3}\vee \overline{x}_{2}) \wedge  (x_{3}\vee x_{2}) \wedge
(\overline{x}_{3}\vee x_{2}) \wedge  (\overline{x}_{1}\vee
\overline{x}_{0}) \wedge  \text{ } (x_{1}\vee x_{0}).$ It has a SSAT$\left(2,3\right)$ (3 first clauses) with solutions $x_{3}=1$ and $x_{2}=1.$ It has a SSAT$\left( 2,2\right)
$ (2 last clauses) with solutions
$\{(x_{1}, x_{0}) \, | \ (0,1) \vee (1,  0) \}.$ The the resulting satisfactory assignations of $\varphi_5$ are $\left\{ \left( 1,1\right) \right\}
\times \left\{ \left( 0,1\right) ,\left( 1,0\right) \right\}$ $=$ $\{
(x_{3}, x_{2},x_{1}, x_{0}) \, | \, (1, 1, 0,  1) \vee (1, 1, 1,
0)\}.$

For 2) $\varphi_6=$SAT$\left( 4,5\right)$ $=$
 $(x_{3}\vee \overline{x}_{2}) \wedge
 (\overline{x}_{3}\vee \overline{x}_{2})
\wedge  (\overline{x}_{3}\vee x_{2}) \wedge  (x_{2}\vee
\overline{x}_{1}\vee x_{0}) \wedge  (x_{2}\vee  x_{1}\vee
\overline{x}_{0}).$ It has a SSAT$\left( 2,3\right) $ (3 first clauses) with solutions $\{ (x_{3}, x_{2})
\,|\, (0, 0) \} .$ It has a SSAT$\left( 3,2\right) $ (2 last clauses) with solutions $\{ (x_{2},
x_{1},  x_{0}) \,|\, (1,  0,   0) \vee (1, 0,  1) \vee (1, 1, 0) \vee
(1,  1, 1) \vee (0, 0,  0) \vee (0,  1, 1) \}.$ Then $\varphi_6
\in$\LSAT, because there are satisfactory assignations for the common value 0 of the variable $x_{2}$. The satisfactory assignations of $\varphi_6$ are $\{ (x_{3}, x_{2}, x_{1}, x_{0}) \,| \, (0,  0,
0, 0) \vee (0,  0,  1, 1) \}.$

For 3) $\varphi_7=$SAT$\left( 4,7\right)$ $=$ 
$(x_{3}\vee \overline{x}_{2}) \wedge  (\overline{x}_{3}\vee
\overline{x}_{2}) \wedge (\overline{x}_{3}\vee x_{2}) \wedge
(x_{2}\vee \overline{x}_{1}\vee \overline{x}_{0}) \wedge
(x_{2}\vee \overline{x}_{1}\vee x_{0}) \wedge (x_{2}\vee
x_{1}\vee \overline{x}_{0})
\wedge (x_{2}\vee x_{1}\vee x_{0})%
.$ It has a SSAT$\left( 2,3\right) $ (3 first clauses) with solutions $\{ (x_{3},  x_{2}) \, | \,
(0,  0) \} .$ It has a SSAT$\left( 3,4\right) $ (4 last clauses) with solutions  $\{ (x_{2},
x_{1}, x_{0}) \,| \, (1,  0,  0) \vee (1, 0,  1) \vee (1,  1, 0) \vee (
1, 1, 1) \}.$ Then $\varphi_7 \notin$\LSAT, because there is not a common value for the variable $x_{2}$.


\begin{proposition}
~\label{prop:CompSATtwo} The random search algorithm~\ref{alg:SAT_two} is computable and its iterations are limited by $2^{n-p}$ when it uses $2^p$ independent processors for the algorithm~\ref{alg:test Cand}.

\begin{proof}
The algorithm has steps that they are easily to mimic by appropriate Turing machines for generation of binary numbers, integer operations and identification. Therefore, it is computable. 

The execution in parallel of the algorithm~\ref{alg:test Cand} by $2^p$ independent processors has the effect to explore $\Sigma^n$ by splitting it into $\Sigma^p$ and one random string from $\Sigma^{n-p}.$ Therefore, the iterations for reviewing all possible satisfactory candidates $\Sigma^n$ is limited by $\frac{2^n}{2^p}=2^{n-p}.$

In any of the independent processors, the condition $\varphi(c)==1$ has the effect to interrupt and stop all, because $c$ becomes a witness, i.e., a satisfactory assignation for $\varphi$. Otherwise, the failed candidate is sent to the algorithm~\ref{alg:UpdtCand}. This does not increase the complexity, it is a simple sending of a message without any communication protocol.
\end{proof}
\end{proposition}

It is worth to note that the randomness of selecting one random candidate from the interval $[0,2^{n-p}-1]$ is losing by combining  all string of $\Sigma^p$ with it. Also, the upper limit $2^{n-p}$ is similar to $2^{n}$ when $p \ll n$ for a SAT with a huge number of variables $n$.

\begin{proposition}
~\label{prop:CompUpdCand} The algorithm~\ref{alg:UpdtCand} for tracking the failed candidates, which it sequentially look for a satisfactory candidate is  computable and its iterations are limited by $2^{n}.$
\begin{proof}

The algorithm hast two main parts one for receiving the failed candidates and one for a sequential search of the viable candidates. Each of these sections can be mimic by appropriate Turing machines for for generation of binary numbers, integer operations and identification. The reception of the failed candidates is a FIFO queue by using a list. In any time, such list is finite because $\Sigma^n$ is finite and the messages from the other algorithms are limited by it. The algorithm executes in sequential way one after the other in a loop limited by the number of possible candidates $2^n$. The effect of keeping track of the failed candidates is to decrease the number of viable candidates to $0$. And this will happen in a finite time because even without any failed candidate message, the algorithm tests all the $2^n$ possible candidates of the circular list.

The algorithm stops all when a viable candidate $c$ $\in$ $\Sigma^n$ fulfills  $\varphi(c) \equiv 1$, i.e., $c$ is a satisfactory assignation for $\varphi$. Otherwise, it mark $c$ as failed candidate and update the circular list cand\_stat.
Therefore, it is computable and limited by $2^n$ iterations.
\end{proof}
\end{proposition}

Assuming that the algorithms runs with similar time slot, the algorithm~\ref{alg:UpdtCand} decreases by around $2^{p+1}$ the number of the $2^n$ possible candidates in each iteration.

\begin{proposition}
~\label{prop:Comp} The parallel algorithm~\ref{alg:Parallel} is computable and its complexity is limited by $2^n$ iterations or $\varphi$'s number of clauses plus the time of the operation $\times \, \theta$.

\begin{proof}
The computability follows from the propositions~\ref{alg:UpdtCand},~\ref{alg:SAT_one} and~\ref{alg:SAT_two}.

In fact, for any SAT problem, the parallel algorithm~\ref{alg:Parallel} only has two unique answers: 1) $\varphi\in $ \LSAT, because it exists $x$ $\in$ $\Sigma$, such that $\varphi(x)\equiv 1$, or 2) $\varphi \notin$ \LSAT, because it does not exist $x$ $\in$ $\Sigma^n$ such it satisfies $\varphi(\cdot)$. This happen by a short termination or by reviewing $\Sigma^n$ or by reading all $\varphi$'s clauses and the conciliation of the solutions of SSAT for determining one satisfactory assignation or none.

The iterations of the algorithms~\ref{alg:SAT_two} and~\ref{alg:UpdtCand} are bounded by the number of candidates $2^n$.

On the other hand, the algorithm~\ref{alg:SAT_one} need to read all $\varphi$'s clauses and performs $\times\,\Theta$ operation for conciliation of the solutions of SSAT.

Therefore the parallel algorithm~\ref{alg:Parallel} is computable and it finishes at after $2^n$ iterations or after it reads the $\varphi$'s clauses plus the time of the $\times \, \theta$ operation.

\end{proof}
\end{proposition}

The probability function with respect to the number of $s$ satisfactory assignations and $k$ failed trials is
$$ 
\Prob(s,k) = \frac{s}{2^n-k}. 
 $$

 Therefore, for finding a unique solution after testing $k$ random different failed candidates decays in exponential way: $$\Prob(1,k) = \frac{1}{2^n-k}.$$

\begin{figure}[H]
\begin{center}
\psfig{figure=\IMAGESPATH/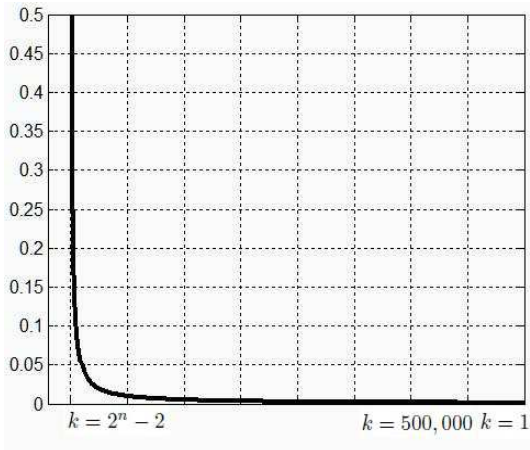,height=60mm,width=70mm}
\caption{Behavior of the function $\Prob(s,k)$ where $s$ is the number of satisfactory assignations and $k$ is the number of trials}
~\label{fig:probDecy}
\end{center}
\end{figure}

The exponential divisor $2^n$ causes a rapidly decay as it depicted in fig.~\ref{fig:probDecy} where $k=2^n-2,$ $k=500,000$ and $k=1$. When $n \gg 0$ is a big number, only after testing a huge number $k=2^n-2$ of the different candidates the probability grows to $0.5$. Meanwhile, for a reasonable number $k \ll 2^n$ of the different candidates, the probability remains insignificant, i.e., $$\Prob(1,k) \approx \frac{1}{2^n} \approx 0.$$

\begin{proposition}
~\label{prop:prob_alg_sol} For any SAT problem:
$$\Prob_{\ref{alg:UpdtCand}}(k,s)
 \leq \Prob_{\ref{alg:SAT_one}}(k,s)
 \leq \Prob_{\ref{alg:SAT_two}}(k,s)$$
where $\Prob_{i}(k,s)$ is the probability of finding a solution by the $i$ algorithm after $k$ failed iterations, $i=~\ref{alg:UpdtCand},~\ref{alg:SAT_one}, and ~\ref{alg:SAT_two},$ and $s$ is the number of satisfactory assignations.
\begin{proof}
The probability for the algorithm~\ref{alg:SAT_one} is:
$$ 
\Prob(s,k) = \frac{s}{2^n-k}. 
 $$

The probability for the algorithm~\ref{alg:SAT_two} using $2^p$ independent processors is:
$$
\Prob(s,k) = \frac{s}{2^n-k2^{p}}. 
 $$

And finally, the probability for the algorithm~\ref{alg:UpdtCand} is approximately (the estimation is affected by losing messages and repeated failed candidates form the algorithms~\ref{alg:SAT_one} y ~\ref{alg:SAT_two}):
$$ 
\Prob(s,k) \approx \frac{s}{2^n-k(2^{p+1})}. 
 $$
 \end{proof}
\end{proposition}

When the number of clauses is a small number $m$, the probability for finding a satisfactory assignation is high, because it means that there are around $2^n-m$ candidates not blocked by the small number of clauses $m$. Under this situation the algorithm~\ref{alg:SAT_one} could find the solution before the algorithms~\ref{alg:SAT_two} and ~\ref{alg:UpdtCand} not because its probability but the $\varphi$'s small number of clauses. The previous proposition in fact states that the algorithm~\ref{alg:UpdtCand} will be solve the problems more frequently of the other algorithms under the assumption that they send their failed candidates to it.

I called an extreme SAT problem, a SAT problem with a huge number of variables $n$, where its clauses uses at most $n$ variables, the clauses could be repeated, and in disorder, but the most important characteristic is that an extreme problem has one or none solution, i.e., one satisfactory binary number or none. It means in the case of one solution that the probability to guess the solution is $\frac{1}{2^n}$, and for none solution is 0. Figure~\ref{fig:probDecy} depicts this behavior.
Therefore, for $m$ extreme problems SAT, the expected value for finding the solution is almost 0.

The next proposition analizases the effect of $2^k$ independent processors for the parallel execution of the algorithm~\ref{alg:UpdtCand} for an extreme SAT.
\begin{proposition}
~\label{prop:parallel_randseaarch} For any extreme SAT with a huge number of variables $n$ ($n\gg 0$). Then the algoritmo~\ref{alg:Parallel} and particularly, the algorithms ~\ref{alg:UpdtCand}, and ~\ref{alg:SAT_two} do not improve the efficiency.
\begin{proof}

The parallel algorithm~\ref{alg:Parallel} executes in parallel the
algorithms~\ref{alg:SAT_one} and the random search algorithm~\ref{alg:SAT_two}.

Without loss of generality, it is possible to have $2^p$ processors for the execution of the algorithm~\ref{alg:test Cand}.
With $2^p$ a reasonable numbers of processors, the testing at the same time of the $2^k$ random unique candidates by the algorithm~\ref{alg:test Cand}, which it is called in parallel by the random search algorithm~\ref{alg:SAT_two}, takes $2^{n-p}$ iterations. 
But for an extreme SAT, $n \gg 0$ is a huge number and the $2^p$ processors is small and reasonable, then $n \gg p$, and the time of the iterations for testing the candidates is $2^n$ because $n \approx {n-p}.$

Even the algorithm~\ref{alg:UpdtCand} does not improve the efficiency. The proposition~\ref{prop:prob_alg_sol} states that it is the algorithm with high probability to solve SAT. Its number of viable candidates decrease by approximately $2^{p+1}$ in each iteration. Therefore, after a reasonable number of iterations $k$, the number of viable candidates is approximately $2^n-k(2^{p+1})$. Considering that $k \ll n$ and $p\ll n$ are reasonable numbers comparing to a huge number $n$, 
 $2^n-k(2^{p+1})\approx 2^n.$

No matters the algorithms~\ref{alg:UpdtCand},~\ref{alg:SAT_one} and~\ref{alg:SAT_two}, the probability for solving SAT is approximately
$\frac{s}{2^n} \approx 0$ for $k$ and $2^p$ reasonable numbers, and $n$ is a huge number $n$, and $|s| \leq 1$.

In the case $s=0$ the only way to know that there is not satisfactory assignation is by reviewing all $\Sigma^n$, therefore there is way to improve the efficiency.

\end{proof}
\end{proposition}

For an extrema SAT, $n \gg 0$ and with one satisfactory assignation or none, it is almost impossible to determine the unique satisfactory assignation in efficient time, but it is worst when there is not solution, the probability always is equals to zero (see figure~\ref{fig:probDecy}) and it is necessary to verify that none of the candidates of the $2^n$ candidates of $\Sigma^n$ are satisfactory assignations. Other alternative is the quantum computational model~\cite{Zhang2002,arXiv:Barron2015b}.

\section{Results}~\label{sc:results}

Any of the formulations $r,1$-SAT o $r,2$-SAT o $r,r$-SAT is solved by the algorithm~\ref{alg:Parallel} in linear time with respect to the number of the $\varphi$'clauses. Therefore its efficiency is less or equal than the state of the art algorithms for SAT
~\cite{Pudlak1998,Zhang:2001:ECD:603095.603153,Zhang2002,TOVEY198485}.

The lower complexity is linear, and it is worth to analice parallel processing. It could reduce the complexity, however this is not the case for the parallel algorithm as it is depicted in the previous section for an extreme SAT problem.

\begin{proposition}~\label{prop:NotEfiAlgSAT} 
There is no efficient algorithm to solve an extreme SAT problem.
\begin{proof} 

Suppose there is an efficient algorithm to solve any SAT problem. Obviously, an extreme SAT problem must be solved by such an algorithm.

Two people are selected, the number one person defines the extreme SAT problem with the freedom to decide one or no solution. The second person has a powerful computer and has the efficient algorithm.

How long will it take for the second person to solve a series of $M$ extreme SAT problems? He has the efficient algorithm, so he has to give the two possible answers in a short time. The assignment of satisfactory values or that there is no solution.

If the efficient algorithm fails, no matter the time, it is useless.

On the other hand, for reasons of argument, suppose that the second person with his powerful computer and the efficient algorithm gets the solution of a series of $M$ SAT extreme problems in a reasonable time. The value of $M$ is a reasonable value for human perception but with 
$M$ $\ll$ $2^n$. There is a contradiction. A succession of $M$ consecutive successes for solving extreme SAT problems means that the first person does not freely and arbitrarily decide extreme SAT problems. The expected value for honestly solving a series of $M$ extreme SAT problems is 0.
\end{proof}
\end{proposition}

With colleagues and students, I proposed an alternative argument. A lottery company defines its winning ticket for an extreme SAT problem. When extreme SAT problems have a unique solution, no one complains. The winner's ticket satisfies everyone because, verification is done in efficient time, and there is a witness, the winning ticket. But, when the extreme SAT problem has no solution. There is no winner. So it's hard to accept, because with a large number $n$ of variables, it takes a lot of time to verify that none of the $2^n$ tickets satisfies the extreme SAT problem. And without this verification, the result that there is no winner rests on the honesty of the lottery company.

Another aspect is the linearity of the algorithm with respect to the SAT's number of clauses. Of course, it is linear, but no person or computer can solved a SAT problem without reviewing the SAT's clauses. On the other hand, how can an extreme SAT problem be constructed if the number of its clauses is exponential and about or greater than $2^n$. Well, here I have three positions: 1) it is a hypothesis, a valid theoretical assumption; 2) the advances in the research of nanotechnology, clusters of molecules and crystal structures will soon provide capable and complex electronic circuits as extreme SAT problems, and 3) a practical experiment can be performed using the logical equivalence between CNF and DNF (Normal Disjunctive Form). An extreme SAT problem in DNF is the empty set of clauses or a binary number as a DNF clause. For example, if the only solution to an extreme SAT problem is $001$, then the DNF clause is $(\overline{x}_2 \wedge \overline{x}_1 \wedge x_0)$. It is possible to simulate an extreme SAT problem with a random permutation (with computer's pseudo random numbers) on the fly like the algorithm~\ref{alg:SAT_two}. For the second, a circuit will work, but reviewing all SAT's clauses is the only way to make sure that it works under its design's specification. For the latter, I did a lottery experiment for an extreme SAT problem with colleagues and students, persistent people give up after weeks of asking me, is this the number? Despite the fact that I offer them a million dollars as a motivation to build their algorithm and defeat me.

\section{Conclusions and future work}~\label{sc:conclusions and future work}

The modified parallel algorithm of this article explores the use of $2^p$ independent processors  for parallel execution to improve the parallel algorithm without algebra in~\cite{arXiv:Barron2016b,comtel2016:Barron2016a}. The main results are 1) the linear complexity of the modified parallel algorithm~\ref{alg:Parallel} and 2) its implication that there is not an efficient algorithm for solving an Extreme SAT problems: proposition~\ref{prop:NotEfiAlgSAT}.

There are open problems $r,s$-SAT as by example the conjecture 2.5 in~\cite{TOVEY198485}. But nevertheless, under the assumptions of enough memory and time any arbitrary formulation of SAT can be solved by the algorithm~\ref{alg:Parallel}.

It is viable to extend the algorithm~\ref{alg:Parallel} for including formulation of SAT using the logical operators $\Rightarrow$, $\Leftrightarrow$, CNF, DNF, and nested parenthesis. More details are in~\cite{arXiv:Barron2015b,arXiv:Barron2016,arXiv:Barron2016b}.


\end{multicols}

\end{document}